\documentclass[amsfonts,amsmath,amssymb,prb,showpacs,twocolumn]{revtex4}
\usepackage{graphicx}
\usepackage{color}
\usepackage{bm}
\usepackage{bbm}
\usepackage{dsfont}
\usepackage{mathbbol}
\usepackage{multirow}

\newcommand{\e}{\begin{equation}}
\newcommand{\ee}{\end{equation}}

\newcommand{\ea}{\begin{eqnarray}}
\newcommand{\eea}{\end{eqnarray}}

\newcommand{\vectornorm}[1]{\left|\left|#1\right|\right|}
\newcommand{\ket}[1]{\left. | #1 \right\rangle }
\newcommand{\bra}[1]{\left\langle #1 |\right. }
\newcommand{\einheit}[1]{\hspace{1mm}\mbox{#1}}
\newsavebox{\mysquare}
\savebox{\mysquare}{\textcolor{black}{\rule{2.5mm}{2.5mm}}}
\usepackage{float}

\begin{document}
\title{Time--optimal performance of Josephson charge qubits: A process tomography approach}
\author{Robert Roloff}
\email{robert.roloff@uni-graz.at}
\author{Walter P\"{o}tz}
\email{walter.poetz@uni-graz.at}
\affiliation{Fachbereich Theoretische Physik, Institut f\"{u}r Physik, Karl--Franzens Universit\"{a}t Graz, Universit\"{a}tsplatz 5, 8010 Graz, Austria}
\date{\today}

\begin{abstract}
A process tomography based optimization scheme for open quantum systems is used to determine the performance limits of Josephson charge qubits within current experimental means.  The qubit is modeled microscopically as an open quantum system taking into account state leakage, as well as environment--induced dephasing based on experimental noise spectra. Within time--optimal control theory, we show that the competing requirements for suppression of state leakage and dephasing can be met by an external control of the effective qubit--environment interaction, yielding minimal gate fidelity losses of around  $\Delta F \approx 10^{-3}$ under typical experimental conditions.
\end{abstract}

\maketitle
\section{Introduction}
Within the circuit model of quantum computing, the fundamental building block of a quantum computer is the \textit{quantum bit} or \textit{qubit}. 
Our ability to precisely execute unitary operations within qubits is an essential prerequisite for the implementation of quantum algorithms and for harnessing the full computational power of more complex quantum system.
Among the various proposals for a physical realization of a qubit, specially designed superconducting circuits have been identified as promising candidates.~\cite{Shnir97,Mak01}  Their potential is mostly founded on the high technological standards by which SQUIDS can be fabricated and controlled, as well as their promise regarding array scalability.~\cite{You03}  Several types of Josephson--junction--based qubit designs, such as charge, phase, and flux qubits, have been proposed and explored in the laboratory.~\cite{Nak99,Martinis02,Orl99,Clarke08}   
Conditional gate operations have already been performed experimentally.~\cite{Yam03,Plant07}

While preliminary results indeed look promising, considerable improvement in gate performance will be necessary to make these structures  
useful in larger arrays of quantum gates. Similar to all solid--state--based qubits, Josephson qubits suffer from two main shortcomings: 
they are quantum two--level systems only within approximation and there is non--negligible coupling to the environment.  
The former may result in leakage to non--computational basis states which affects the fidelity of quantum gates.~\cite{Fazio99,Monta07,Spoerl07,Reben09,Motzoi09}
The latter results in an unwanted population relaxation and  destruction of state superpositions, both being detrimental for quantum computation.~\cite{Asta04,Ithier05}

In this work we study Josephson charge qubits and show how the conflicting requirements for suppression of both state leakage and decoherence can be accomplished within optimal control theory. In the first part, we outline the model for the Josephson charge qubit including leakage and environmental interaction on which we base our study. Then, we formulate a cost functional for state--independent optimization of open quantum systems based on the Kraus representation and process tomography. This method can be readily applied to any open quantum system, including other qubit implementations as well as multi--qubit gates and is not necessarily restricted to solid--state realizations. In the remainder of the paper, we demonstrate the approach for the Hadamard gate implementation.\par

\section{Josephson Charge Qubit}
\label{JCQ}
\begin{figure}[h]
\centerline{\includegraphics[width=5cm]{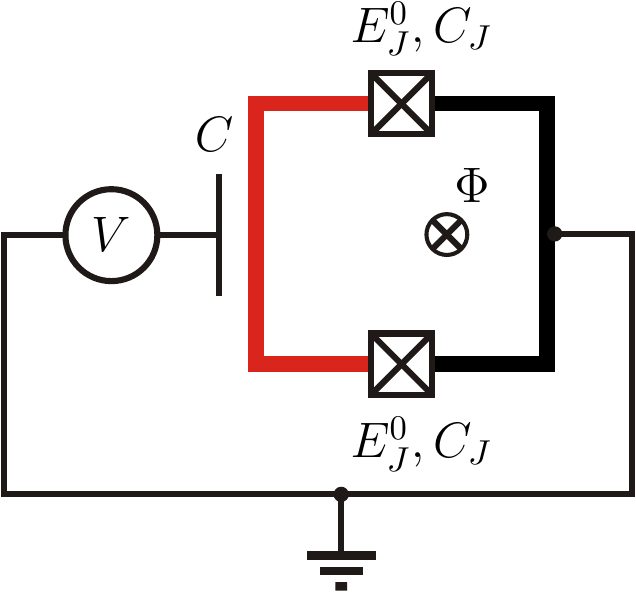}}
\caption{\label{qubit1}(Color online) Superconducting charge qubit with 2 Josephson junctions. The two junctions are realized by small insulating layers which divide a superconducting ring into a small island (red/gray) and a reservoir (thick black).~\cite{Mak01} The qubit is driven by an external flux $\Phi$ and a gate voltage $V$. The voltage is coupled capacitively ($C$) to the island. The Josephson junctions are characterized by a capacitance $C_J$ and the Josephson coupling energy $E_J^0$.}
\end{figure}

The basic structure of a  Josephson charge qubit  is shown in Fig.~\ref{qubit1}.~\cite{Mak99,Nak99,Mak01}  The  characteristic energy scales are the charging energy $E_C$ of the superconducting island, the Josephson coupling energy $E_J^0$ and the superconducting energy gap $\Delta$.
If $\Delta$ is the largest energy, the problem can be reduced to a situation where no quasiparticle excitation is found on the island and only Cooper pairs can tunnel through the Josephson junctions. Cooper--pair relaxation on the island, as well as quasiparticles tunneling from the reservoir onto the island, contribute, in principle, to relaxation and decoherence. The latter takes place at time scales of the order $\left( T_2^{qp}\right)^{-1} \sim \frac{g_T \delta_r}{2 \pi \hbar}N_{qp}$, where $g_T$ denotes the conductance of the Josephson junctions (in units of $e^2/h$), $\delta_r$ is the quasiparticle level spacing in the reservoir and $N_{qp}$ is the number of quasiparticles.~\cite{Lutchyn05,Lutchyn06} For typical charge--qubit architectures, $T_2^{qp}$ is of the order of $10^{-3}\einheit{s}$ to $10^{-5}\einheit{s}$ and, thus, on a time scale much larger than we are investigating in the present work. Cooper--pair relaxation within the island is even weaker and may be neglected, too. 

Within the charge basis $\{\ket{n}\}$, the Hamiltonian of the system,
\begin{equation}\label{Hamiltonian}
H_S=4E_C\left[ n - n_c(V)\right]^2-2E_J^0\cos{\left( \pi \frac{\Phi}{\Phi_0}\right) }\cos{\Theta},
\end{equation}
may be written as,
\ea
H_S&=&\sum\limits_{n}{ \left\lbrace  4 E_C\left[ n-n_c(V)\right] ^2\ket{n}\bra{n}+\right.} \nonumber \\
& & \left. E_J(\Phi)\left( \ket{n+1}\bra{n}+\ket{n}\bra{n+1}\right) \vphantom{\left[ n-n_c(V)\right] ^2}\right\rbrace ,
\eea
with $\Theta$ (the phase difference across a Josephson junction) being the conjugate variable to the number $n$ of additional Cooper pairs on the island.~\cite{Mak01}
\begin{equation}\label{EJPhi}
E_J(\Phi)=-E_J^0\cos{\left( \pi \Phi / \Phi_0\right) }
\end{equation} and $n_c(V)=V C/(2e)$ contain two independent physical control fields, $V=V_g+V_p$ and $\Phi$, where $V_g$ and $V_p$ denote a dc and a pulse gate voltage, respectively. $\Phi$ is the external magnetic flux. The latter can be used to tune $E_J(\Phi)$ between $-E_J^0$ and~$0$.~\cite{Barone82} By proper choice of the dc gate voltage $V_g$, one can set the qubit ``working point'' to the charge degeneracy point, where the qubit is insensitive to charge fluctuations up to the first order. Using, next to the computational basis $\{\ket{0},\ket{1}\}$, two adjacent leakage states $\ket{-1}$ and $\ket{2}$, the effective ``leaky qubit" Hamiltonian at the charge degeneracy point reads,
\begin{eqnarray}\label{Hmatrix}
H_S&=&\begin{bmatrix}
8 E_C & E_J(\Phi) & 0 & 0 \\
E_J(\Phi) &  0 & E_J(\Phi) & 0 \\
0 & E_J(\Phi) & 0 & E_J(\Phi) \\
0 & 0 & E_J(\Phi) & 8 E_C
\end{bmatrix}+4E_C n_p(t) X, \nonumber \\
X&=& \begin{bmatrix}
3 & 0 & 0 & 0 \\
0 &  1 & 0 & 0 \\
0 & 0 & -1 & 0 \\
0 & 0 & 0 & -3
\end{bmatrix}, \quad n_p=n_p(V_p).
\end{eqnarray}
For our computation we consider a typical charge qubit with energies $E_C = 150\einheit{$\mu$eV}$ and $E_J^0 = 35\einheit{$\mu$eV}$ from the experiment.~\cite{Asta04,Nak99,Nak02} The qubit can be controlled by tuning the external magnetic flux [$\Phi=\Phi(t)$] and the applied pulse gate voltage [$V_p=V_p(t)$].

In the derivation of the Hamiltonian [Eq.~\eqref{Hamiltonian}], the loop--inductance of the superconducting ring has been neglected. For the parameters used in the present work, this approximation is justified if $L \ll \Phi_0^2/\left( 4\pi^2 E_J^0\right) \approx 20 \einheit{nH}$. We also assume that the Josephson coupling energies of the two junctions are identical. However, if $E_J^0 \rightarrow E_J^0+\delta E_J^0$ for one of the junctions, Eq.~\eqref{EJPhi} turns into $E_J(\Phi)=-E_J^0\left| \left( 1+\eta\right)\cos{\left( \pi \frac{\Phi}{\Phi_0}\right) }+ i\eta\sin{\left( \pi \frac{\Phi}{\Phi_0}\right) } \right| $, with $\eta\equiv \frac{\delta E_J^0}{2 E_J^0}$, see e.g. Ref.~\onlinecite{Barone82}. In order to avoid such an asymmetry, it is possible to substitute one of the Josephson junctions in Fig.~\ref{qubit1} by a SQUID, which enables one to tune the Josephson energies to be equal.~\cite{Mak99}

In superconducting qubits the dominant dephasing mechanism (at low temperature) is attributed to noise which changes from $1/f$ to Ohmic behavior at frequencies of typically $k_B T/\hbar$.~\cite{Asta04} The microscopic origin of the $1/f$ noise is not fully understood yet but it is believed that it originates from background charge fluctuations. Ohmic contributions may result from intrinsic sources or from gate lines.~\cite{Asta04,Nak02,Ithier05} We map these fluctuations onto a bath of harmonic oscillators coupling linearly to the qubit.~\cite{Leg87,Makh04} The system is thus modeled by a Hamiltonian of the form,
\begin{eqnarray}
H(t)&=&H_S(t)+H_B+H_{I}, \quad H_B=\sum_k{\hbar \omega_k b_k^\dag b_k},\nonumber \\
H_{I}&=&\hbar X \otimes \sum_k{ g_k \left( b_k^\dag + b_k\right) }\equiv X \otimes \Gamma, \label{hamiltonian}
\end{eqnarray}
where $g_k$ is the effective coupling constant of the spin--boson interaction and $b_k^{\dag}$ and $b_k$ are the bosonic creation and annihilation operators for the mode with frequency $\omega_k$. The dynamics of system, bosonic bath, and the interaction between the two,  respectively,  is  governed by the Hamiltonians $H_S$, $H_B$, and $H_I$, 
whereby the external control field $\epsilon(t)$ to be optimized is contained in $H_S$ only, thus assuming that there is no direct control over $H_B$  and $H_I$.

Apart from correlation functions, noise has to be classified by its amplitude distribution (AD). For the spin--boson model with a thermal bath the AD is normal--distributed (Gaussian noise), which  agrees quite well with experiment, especially at the short time scales  with which we are concerned.~\cite{Asta06}
Guided by the experimental results of Ref.~\onlinecite{Asta04}, we construct  the spectral density by choosing $J(\omega)=J_{1/f}(\omega)+J_{f}(\omega)$ consisting of 
a $1/f$ and an Ohmic contribution,~\cite{Makh04}
\begin{equation}
J_{1/f} =\alpha_{1/f}/\left( 4 k_B T\right), \quad J_{f}\left( \omega \right)  = \alpha_{f}/\left( 2 \hbar\right) \omega.
\label{J}
\end{equation}
For thermal equilibrium, the noise spectrum $S_{\Gamma}\left( \omega \right)$ associated with the operator $\Gamma$ in Eq.~\eqref{hamiltonian} in terms of $J(\omega)$ is then (a tilde denoting the interaction picture),
\begin{eqnarray}
S_{\Gamma}\left( \omega \right)&=& \left\langle \left\lbrace \tilde\Gamma(t), \tilde\Gamma(t') \right\rbrace\right\rangle_\omega =  \int\limits_{-\infty}^{\infty} {d\tau \left\lbrace \tilde\Gamma(t),\tilde\Gamma(t')\right\rbrace e^{i\tau \omega}} \nonumber \\
&=&2\hbar J(\omega)\coth{\left[ \frac{\hbar\omega}{ 2k_B T}\right] }, \quad \tau \equiv t-t'.
\label{S}
\end{eqnarray}
For $\hbar \omega \ll 2 k_B T$ and $\hbar \omega \gg 2 k_B T$, Eqs.~\eqref{J} and~\eqref{S} give the $1/f$ noise spectrum, $S_{1/f}(\omega)\approx \alpha_{1/f}/\omega$, and the Ohmic spectrum, $S_{\omega}(\omega)\approx \alpha_f \omega$, respectively.
The strength of the charge fluctuations (proportional to $\alpha_{1/f}$), has been determined in experiment to saturate at $\sim (10^{-3} e)^2$ for temperatures lower than $200\einheit{mK}$.~\cite{Asta06} The slope of the Ohmic contribution is given in Ref.~\onlinecite{Asta04}. A plot of the specific noise spectrum which we use numerically can be seen in the inset of Fig.~\ref{co}a.

Inspection of Eqs. \eqref{Hmatrix} reveals that simple control strategies for leakage suppression and minimization of decoherence are in conflict with one another. 
The most convenient way to minimize dephasing within the computational subspace would be a large absolute value of the Josephson coupling energy 
$E_J(\Phi)$ because it sets the minimum time ($t_{op}$) needed to perform unitary transformations which incorporate rotations around the $x$--axis of the Bloch sphere (as needed for the Hadamard gate). The faster $t_{op}$, the lesser the effect of decoherence on the gate.
However, in the present superconducting qubit architecture, $E_J(\Phi)$ is also responsible for coupling to the non--computational basis states $\ket{-1}$ and $\ket{2}$. Due to the structure of $X$ in Eq. \eqref{hamiltonian}, the net decoherence rate is enhanced when they participate in the dynamics. Thus, if decoherence--effects are present, the transfer of coherence to the leakage subspace is highly undesirable. In the unitary case, coherence can be transferred back to the computational subspace without loss (given sufficient control). These conflicting requirements, as well as the complexity of an open quantum system, make the design of pulses which maximize the fidelity of the quantum gate a nontrivial task.

\section{State--Independent Optimal Control}
In the context of quantum information processing, coherent control of unitary operations within the qubit is of fundamental interest.~\cite{Palao02} In contrast to the optimization of state--selective transitions, e.g. to maximize the probability of a certain pathway in a chemical reaction,~\cite{Tann85} here we are interested in the optimization of the whole dynamical map. Hence, the qubit should perform a desired unitary operation, irrespective of its initial state. We call this task ``state--independent'' optimal control.

We consider a Hilbert space which is a tensor product of the Hilbert space of the open quantum system  (from now on denoted as \textit{system}) $\mathcal{H}_S$ and the Hilbert space of the environment (which we will call \textit{bath}) $\mathcal{H}_B$, $\mathcal{H}=\mathcal{H}_S \otimes \mathcal{H}_B=\left( \mathcal{H}_{S1} \oplus \mathcal{H}_{S2} \right) \otimes \mathcal{H}_B$.
Within the system we distinguish between computational states on which the gate operation is specified, spanning $\mathcal{H}_{S1}$, and ``leakage levels'' 
spanning $ \mathcal{H}_{S2}$.  
\footnote{The difference between the tensor product and tensor sum structure with respect to optimization has been discussed in Ref.~\onlinecite{Sklarz06}.}

The task of state--independent optimization can be stated as follows. The density matrix of the system, $\rho_S(t)$, evolves in time according to the map, $\rho_S(0) \mapsto \rho_S(t)=\mathcal{E}_t\left\lbrace  \rho_S(0)\right\rbrace,$ for which the superoperator $\mathcal{E}_t$ is functionally dependent upon externally applied control fields $\epsilon(t)$,  \textit{i.e.} $\mathcal{E}_t=\mathcal{E}_t[\epsilon]$. In order to maintain the positivity of $\rho_S(t)$, the map $\mathcal{E}$ has to be \textit{completely positive} and thus can be represented by Kraus operators $K_{m}$,\cite{Niel02,Havel03}
\begin{equation}
\rho_S(t)=\mathcal{E}_t[\epsilon]\left\lbrace  \rho_S(0)\right\rbrace=\sum\limits_{m}{K_{m}[\epsilon](t)\rho_S(0)K^\dag_{m}[\epsilon](t)}, \label{kraus}
\end{equation}
with $K_{m}[\epsilon](t)$ depending on the propagator of the composite system, and hence on $\epsilon(t)$.
We now want to find a control field $\epsilon^*(t)$ for which, at some final time $t_f$, $\mathcal{E}_{t_f}[\epsilon^*]$ approaches the desired mapping $\mathcal{E}_{D}$ as closely as possible.  For quantum gate operations, for example, one would set
\begin{equation}
\mathcal{E}_{D}\left\lbrace .\right\rbrace = U_D(t_f)\left( .\right) U_D^\dag(t_f), \label{opt_unitary}
\end{equation}
where   $U_D(t_f)$ is the unitary operation to be executed within the gate.  In the context of quantum information theory it is useful to formulate a cost functional
within the language of process tomography (see \textit{e.g.} Refs.~\onlinecite{Niel02,Alt03,Poya97,Niel02b}) to define  a measure of how well such a, not necessarily unitary, desired operation has been accomplished. We rewrite the mapping $\mathcal{E}$ by expanding the Kraus operators, $K_m(t)=\sum_n{\alpha_{mn} \bar K_n}$, with $\alpha_{mn}\in\mathbb{C}$ and  $\bar K_{n} \in \mathcal{A}$, where $\mathcal{A}$ denotes a complete basis set of $M \times M$ matrices.~\cite{Alt03}  $M=M_C+M_L$ is the total number of orthonormal basis states, consisting of $M_C$ computational and $M_L$ leakage levels. 
The final state of the quantum system, starting out in $\rho(0)$,  now reads,
\begin{equation}\label{rhotf}
\rho(t_f)=\mathcal{E}_{t_f}\left\lbrace \rho(0) \right\rbrace = \sum_{m,n}{\bar K_{m} \rho(0) \bar K^\dag_{n} \chi_{mn}(t_f)},
\end{equation}
with $\chi_{mn}=\sum_k{\alpha_{km}\alpha_{kn}^*}$. For the example of the Josephson charge qubit described above, one has $M_C=2$, and $M_L=2$ with basis states $\left\lbrace \ket{0}, \ket{1}\right\rbrace $ and $\left\lbrace \ket{-1},\ket{2}\right\rbrace $, respectively. Hence, we are dealing with a $M^2\times M^2=16\times 16$ representation of $\chi$ [see Figs.~\ref{pt}(a) and~\ref{pt}(d)].
The matrix $\chi$ in Eq.~\eqref{rhotf} is usually termed \textit{process tomography matrix} and in order to compute its elements, we choose a fixed set of operators $\left\lbrace \sigma_j \right\rbrace=\mathcal{B}$ (for simplicity we choose $\mathcal{B}=\mathcal{A}$) for which we determine the time--evolution with respect to the mapping $\mathcal{E}$, \textit{i.e.},
\begin{equation}\label{linear1}
\sigma_j(t_f)\equiv\mathcal{E}_{t_f}\left\lbrace \sigma_j \right\rbrace.
\end{equation}
For one--qubit operations with 2 leakage levels or two--qubit gates, the tensor product of Pauli matrices $\tau_i \otimes \tau_j$ with $i,j\in \left\lbrace x,y,z,0\right\rbrace $ and $\tau_0=\openone$ is a convenient choice for $\mathcal{A}$ and $\mathcal{B}$, which we use in present work. In experiment, Eq.~\eqref{linear1} corresponds to preparation in and subsequent time--evolution of suitable different initial states ($\sim \sigma_j$) of the quantum system. Recently, process-tomography methods have been applied to superconducting qubits.~\cite{Neeley08,Chow09}
Because Eq.~\eqref{kraus} is a linear mapping, one can rewrite Eq.~\eqref{linear1} by
\begin{equation}\label{linear2}
\mathcal{E}_{t_f}\left\lbrace \sigma_j \right\rbrace=\sum_k{c_{jk}\sigma_k}.
\end{equation}
The mapping $\mathcal{E}_{t_f}$ is described completely in terms of coefficients $c_{jk}$, whose experimental determination involves quantum state tomography.~\cite{Niel02} The last step in order to calculate the process tomography matrix relates $c_{jk}$ to $\chi_{mn}$. Therefore, we note that, combining Eqs.~\eqref{rhotf}--\eqref{linear2},
\begin{equation}\label{linear3}
\sum_k{c_{jk}\sigma_k}=\sum_{m,n}{\sigma_{m} \sigma_j \sigma_{n} \chi_{mn}(t_f)}.
\end{equation}
To extract the coefficients $c_{jk}$ we apply the scalar product $\left\langle \sigma_i,\sigma_k\right\rangle =\delta_{ik}$ and get from Eqs.~\eqref{linear1}--\eqref{linear3},
\begin{eqnarray}
c_{ji} &=& \left\langle \sigma_i, \sigma_j\left( t_f\right) \right\rangle, \\
&=&\sum_{m,n}{\left\langle \sigma_{i}, \sigma_{m} \sigma_j \sigma_{n}\right\rangle  \chi_{mn}(t_f)}\equiv \sum_{m,n} B_{imjn}\chi_{mn}(t_f). \nonumber
\end{eqnarray}
$\chi_{mn}(t_f)$ can now be calculated by solving the system of linear equations $\left\langle \sigma_i, \sigma_j\left( t_f\right) \right\rangle=\sum_{m,n} B_{imjn}\chi_{mn}(t_f)$, which includes inversion of $B$.

Defining the operator,~\cite{Havel03}
\begin{equation}\label{chitrafo}
\hat \chi=\sum_{m,n}{({\bar K_{n}}^*\otimes \bar K_{m})\chi_{mn}},
\end{equation}
we formulate a simple cost functional,
\begin{equation}
J\equiv \vectornorm{P \hat \chi - \hat \chi^D}^2=\operatorname{tr}\left\lbrace \left[ P \hat \chi - \hat \chi^D\right] \left[ P \hat \chi - \hat \chi^D\right]^\dag \right\rbrace, \label{cost}
\end{equation}
where $P$ denotes the projector onto the $M_C$-dimensional computational Hilbert space and $0\leq J\leq J_{max}=2 M_C^2$.
For the charge qubit example,
\begin{equation}
P_{ij}=\sum_{k=6,7,10,11}{\delta_{i,k}\delta_{j,k}}\quad \mbox{with}\; i,j \in \left\lbrace 1,2,...,M^2=16\right\rbrace. \nonumber
\end{equation}
The transformation given in Eq.~\eqref{chitrafo} corresponds to stacking the columns of the density matrix $\rho_S$ from left to right on top of one another, \textit{i.e.} $\rho_S\rightarrow \operatorname{col}\left( \rho_S\right)$, which yields a $M^2$ dimensional single--column vector, termed $\operatorname{col}\left( \rho_S\right)$.
For the present work, this transformation is of the following form,
\begin{equation}\label{trafo}
\rho_S=
\begin{bmatrix}
\rho_{11} & \rho_{12} & \rho_{13} & \rho_{14} \\
\rho_{21} & \mbox{\fbox{$\rho_{22}$}} & \mbox{\fbox{$\rho_{23}$}} & \rho_{24} \\
\rho_{31} & \mbox{\fbox{$\rho_{32}$}} & \mbox{\fbox{$\rho_{33}$}} & \rho_{34} \\
\rho_{41} & \rho_{42} & \rho_{43} & \rho_{44}
\end{bmatrix}\rightarrow
\operatorname{col}\left( \rho_S\right)=
\begin{bmatrix}
\rho_{11} \\ \rho_{21} \\ \rho_{31} \\ \rho_{41} \\ 
\rho_{12} \\ \mbox{\fbox{$\rho_{22}$}} \\ \mbox{\fbox{$\rho_{32}$}} \\ \rho_{42} \\ 
\rho_{13} \\ \mbox{\fbox{$\rho_{23}$}} \\ \mbox{\fbox{$\rho_{33}$}} \\\rho_{43} \\ 
\rho_{14} \\\rho_{24} \\\rho_{34} \\\rho_{44}
\end{bmatrix},
\end{equation}
where framed density matrix elements belong to the computational subspace. The process tomography matrix $\chi$ changes accordingly, $\chi \rightarrow \hat \chi$, as given in Eq.~\eqref{chitrafo}. For further details see Ref.~\onlinecite{Havel03}. $J$~measures the norm distance between the target operation $\hat\chi^D$ and the actual operation $\hat\chi$ executed at time $t_f$ for control field $\epsilon$.
For the Hadamard gate with 2 leakage levels we set,
\begin{equation}
\hat \chi^D=U_D^* \otimes U_D,\quad \mbox{with}\;U_D=\frac{1}{\sqrt{2}}\begin{bmatrix} 0 & 0 & 0 & 0 \\ 0 & 1 & 1 & 0 \\ 0 & 1 & -1 & 0 \\ 0 & 0 & 0 & 0 \end{bmatrix}. \nonumber
\end{equation}
A similar approach based on a super--operator formalism is given in Ref.~\onlinecite{Wenin08c}.
\section{Qubit Dynamics}
Starting with the von--Neumann equation for the full system dynamics, $\frac{d}{dt} \rho(t) = -\frac{i}{\hbar} \left[ H(t),\rho(t)\right]$, the time evolution of the operators $\tilde \sigma_j(t)$ (with tilde denoting the interaction picture) is computed within a non--Markovian master equation in Born approximation,~\cite{Car02}
\begin{eqnarray}
&&\frac{d}{dt} \tilde \sigma_j(t)= \label{eom}\\
&&-\frac{1}{\hbar^2} \int_0^t{dt'\operatorname{tr_B}\left\lbrace \left[\tilde H_{I}(t),\left[\tilde H_{I}(t'), \tilde \sigma_j(t') \otimes \tilde \rho_B(0) \right] \right] \right\rbrace}, \nonumber
\end{eqnarray}
using a numerical method described in Ref.~\onlinecite{Wenin06} to deal with the kernel, which is non--local with respect to time. $\operatorname{tr_B}$ denotes the partial trace over all bosonic degrees of freedom. For $\tilde \rho_B(0)$ we assume a bath in thermal equilibrium at $T=50\einheit{mK}$.
We also introduce a sharp infrared and a continuous ultraviolet cutoff for the spectral density, \textit{i.e.},
\begin{equation}\label{spectral_sum}
J(\omega) \rightarrow \left( J_{1/f}(\omega) \Theta\left( \omega-\Lambda_{ir}\right) +J_{f}(\omega)\right) e^{-\omega/\Lambda_{uv}}.
\end{equation}
We choose $\Lambda_{ir}=2\pi \times 100\einheit{Hz}$ and $\Lambda_{uv}=2\pi \times 100\einheit{GHz}$ (Ohmic noise in Josephson charge qubits has been measured up to frequencies of $100 \einheit{GHz}$).~\cite{Asta04} The lower bound for the infrared cutoff is determined by the data--acquisition time of the experiment. We have tested our optimal pulses with respect to different infra- and ultraviolet cutoff frequencies, see Figs.~\ref{co}(a) and ~\ref{co}(b). When evaluating the trace of the double commutator in Eq.~\eqref{eom}, one has to calculate bath correlation functions of the form 
\begin{equation}
\left\langle \tilde \Gamma(t) \tilde \Gamma(t') \right\rangle = \operatorname{tr_B}\left\lbrace \tilde \Gamma(t) \tilde \Gamma(t')\tilde \rho_B(0)\right\rbrace, \nonumber
\end{equation}
which can be derived analytically using the spectral density given in Eq.~\eqref{spectral_sum} and~\eqref{J}.\par

\section{Results}
\begin{figure}[h]
\begin{tabular*}{8.6cm}{ll}
\multicolumn{2}{l}{\textbf{Short pulses $\sim 50 \einheit{ps}$}}\\
(a) & (b) \\
\multirow{3}{4.25cm}[2.25cm]{\includegraphics[width=4.25cm]{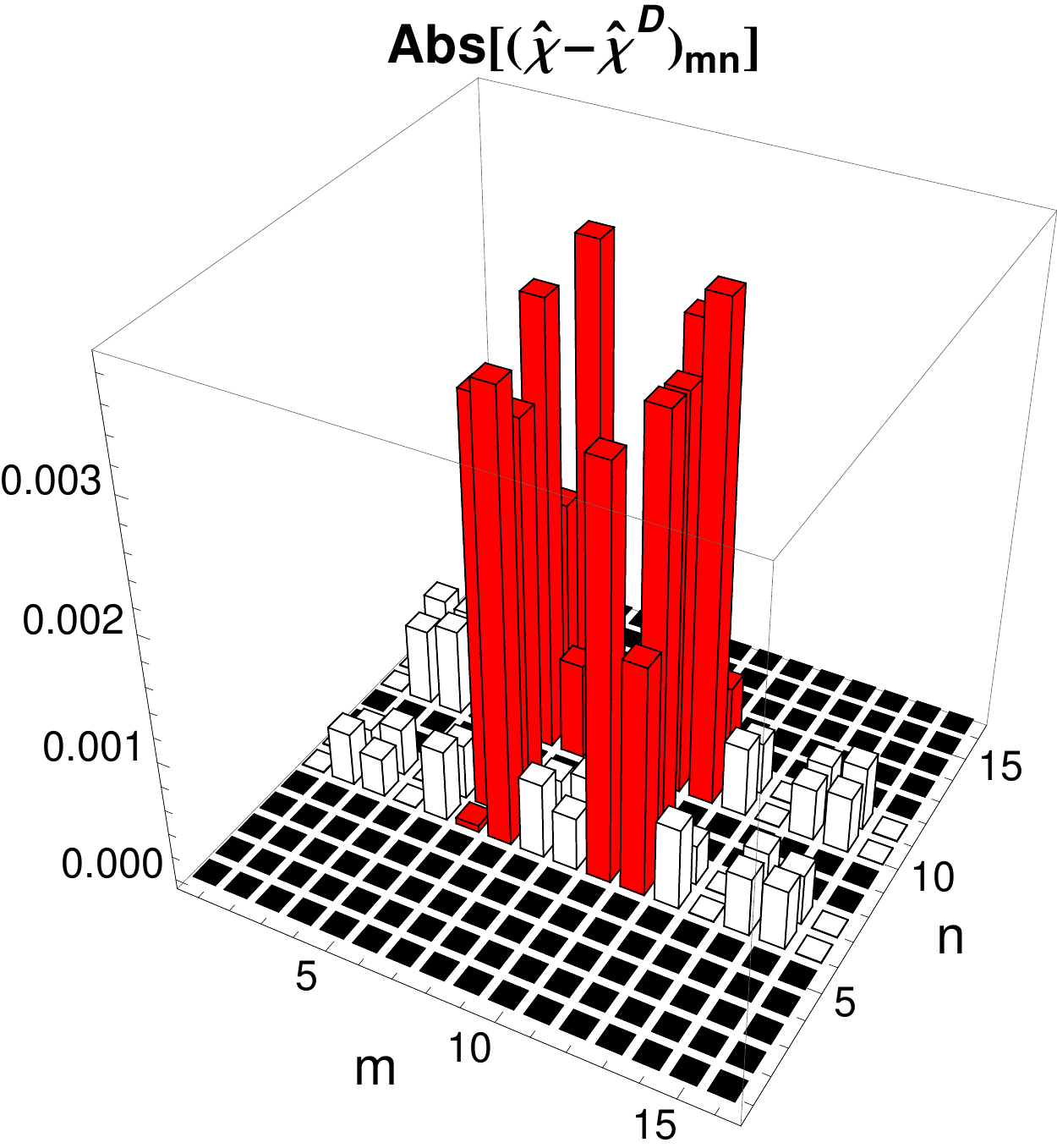}} & \includegraphics[width=4cm]{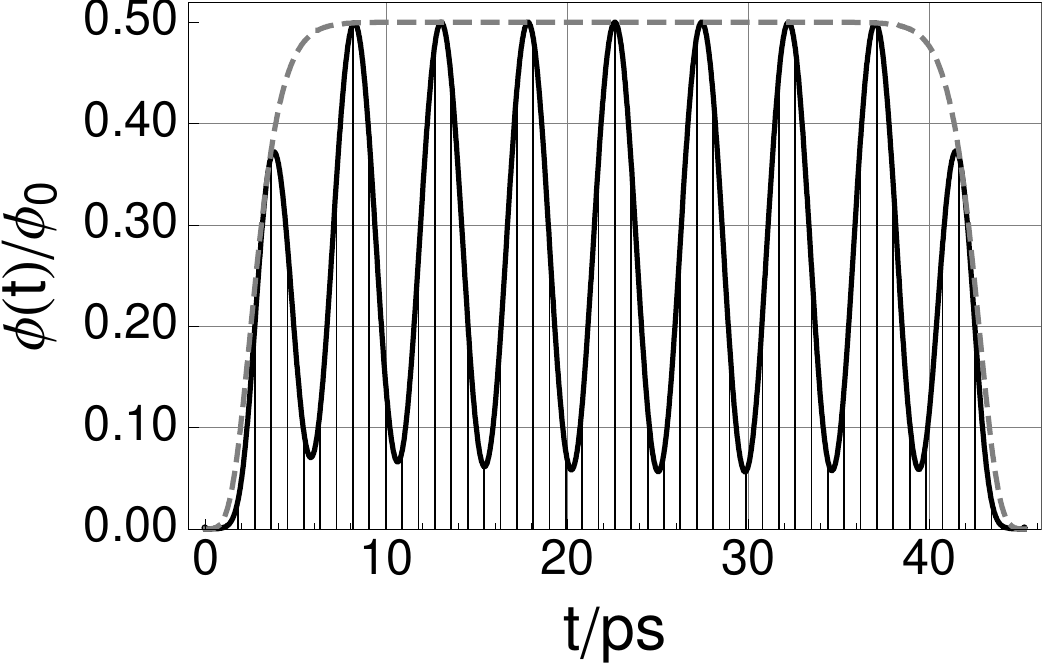} \\
 & (c) \\
 & \includegraphics[width=4cm]{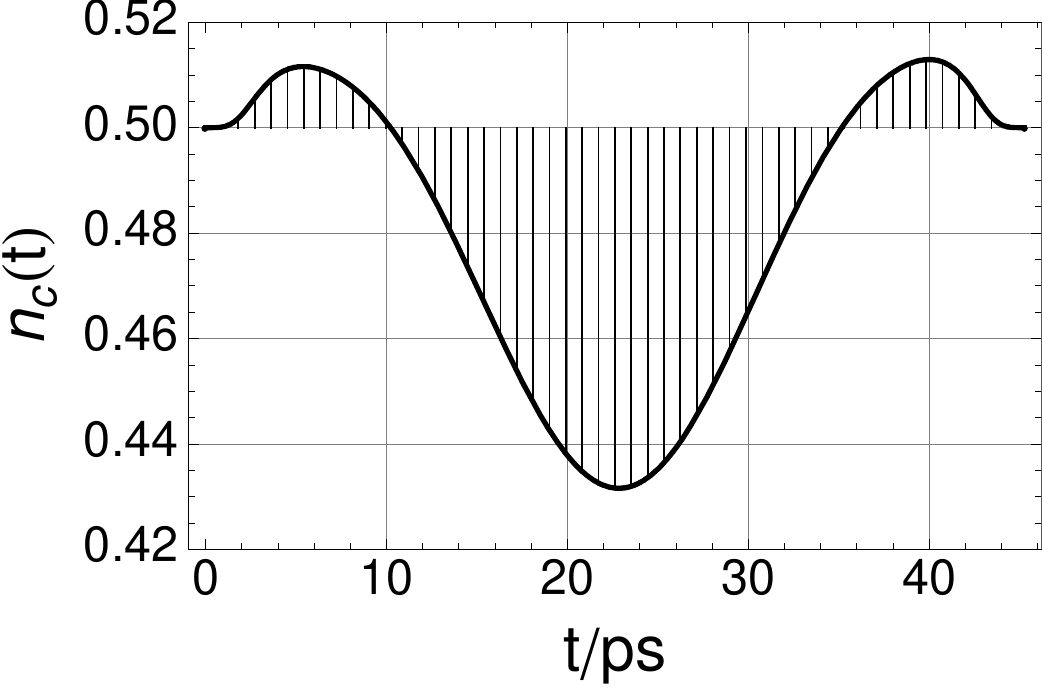}\\
\multicolumn{2}{l}{\textbf{Longer pulses $\sim 200 \einheit{ps}$}}\\
(d) & (e) \\
\multirow{3}{4.25cm}[2.25cm]{\includegraphics[width=4.25cm]{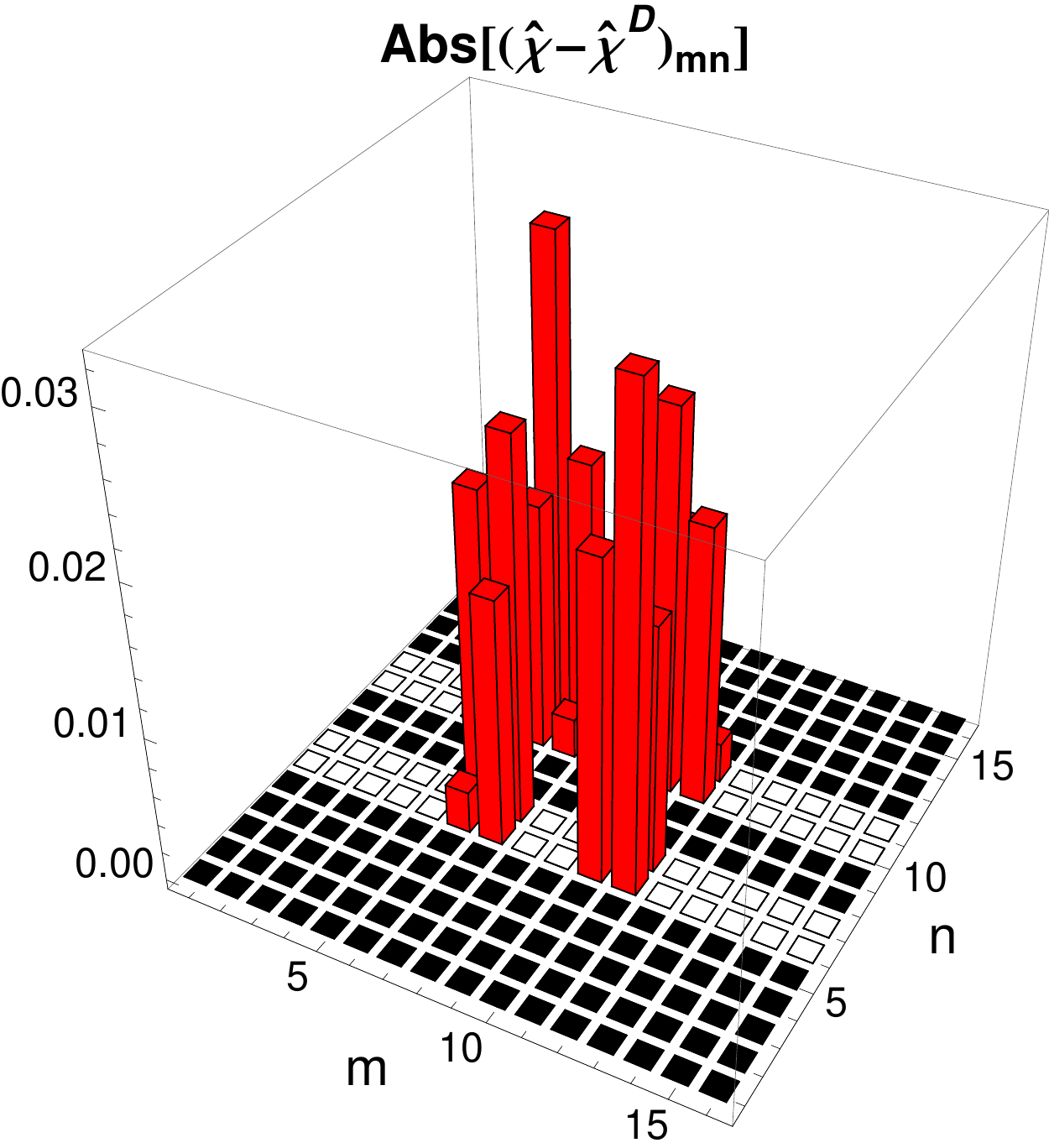}} & \includegraphics[width=4cm]{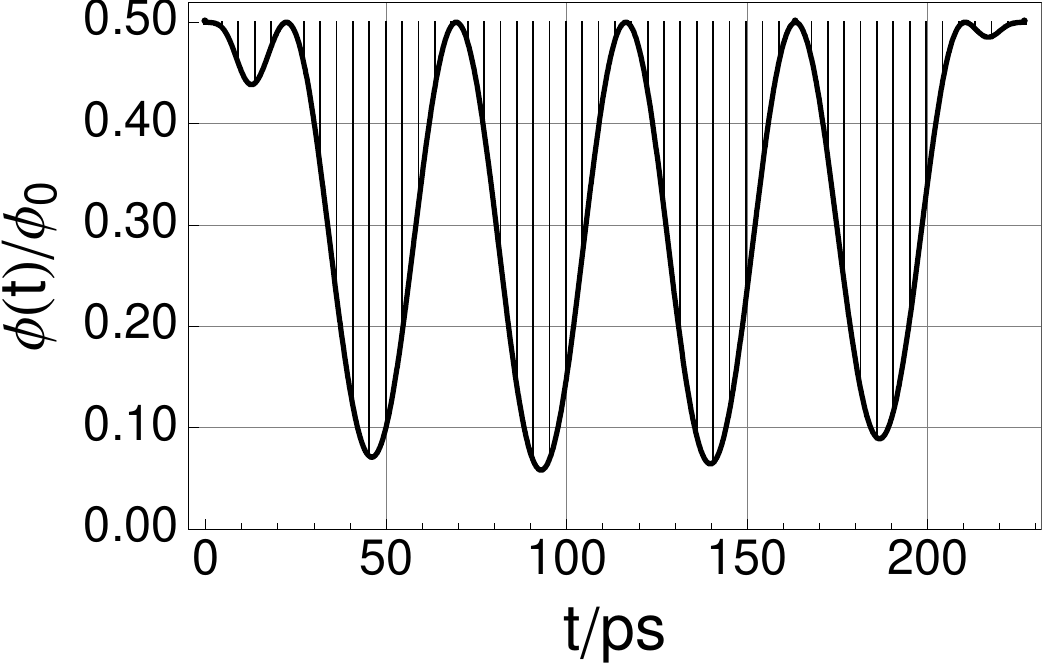} \\
 & (f) \\
 & \includegraphics[width=4cm]{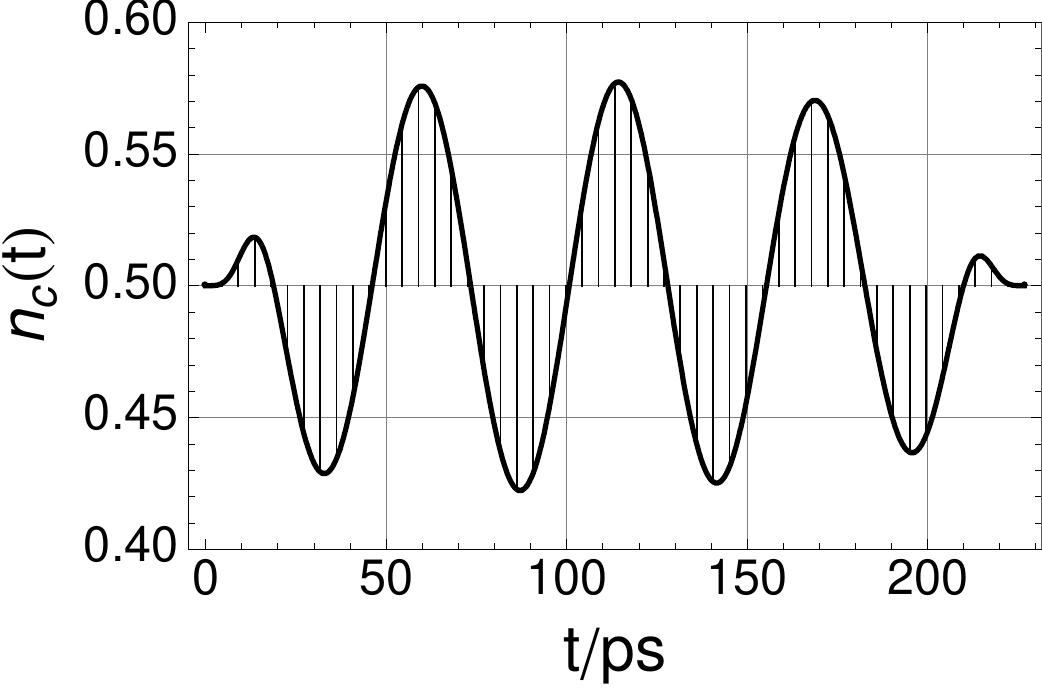}
\end{tabular*}
\caption{\label{pt}(Color online) (a,d) Deviation of the Hadamard process tomography matrix elements $\left( \hat  \chi\right)_{mn}$ for optimized control fields [as given in Figs.~\ref{pt}(b,e) and (c,f)] from desired values $\left( \hat  \chi^D\right)_{mn}$. Black areas denote matrix elements which are irrelevant and, hence, arbitrary. They neither alter the qubit dynamics nor induce transitions out of the qubit subspace, \textit{i.e.} leakage. For simplicity, they are plotted as zero [see also Eq.~\eqref{trafo}]. Red (dark gray) columns show deviations of matrix elements which represent the gate operation within the qubit subspace, whereas white bars display leakage. (b,e) and (c,f) The strength of the optimal control fields lies within realistic values and the optimal solutions have a simple shape.}
\end{figure}

\begin{figure}[h]
\begin{tabular*}{9cm}{ll}
\multicolumn{2}{l}{(a)}\\
\multicolumn{2}{c}{\includegraphics[width=8.6cm]{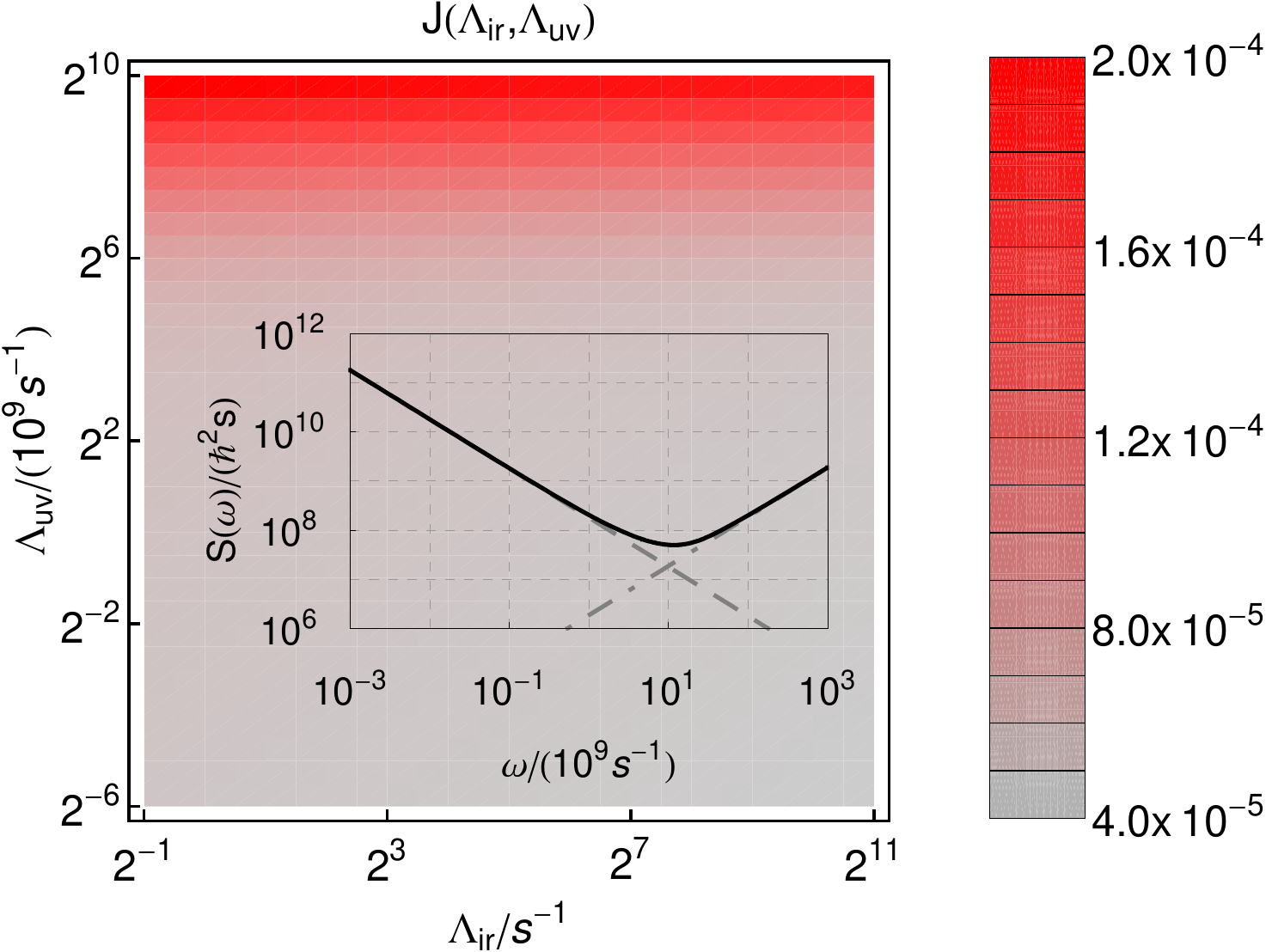}}\\
(b) & (c)\\
\includegraphics[width=4cm]{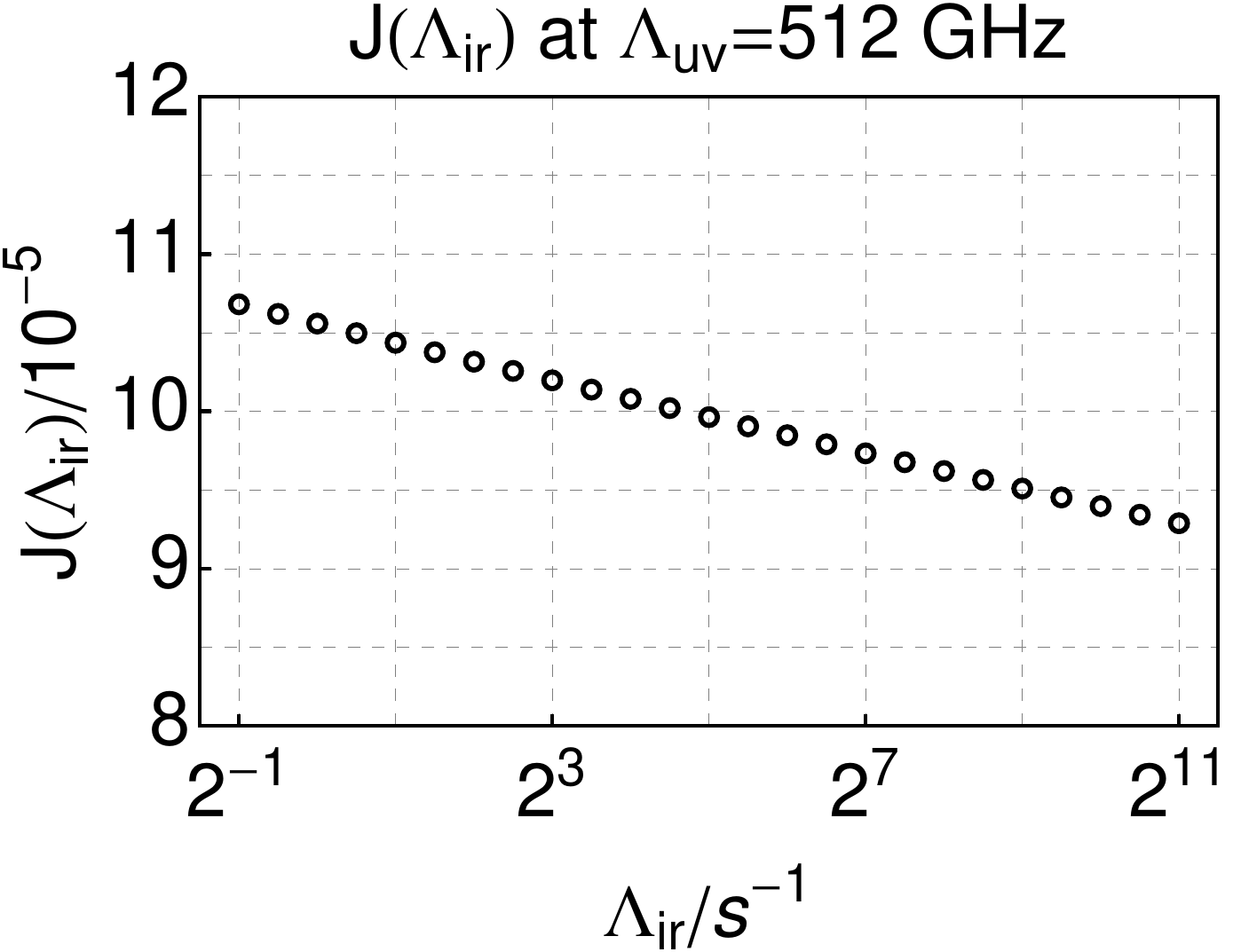}&\includegraphics[width=4cm]{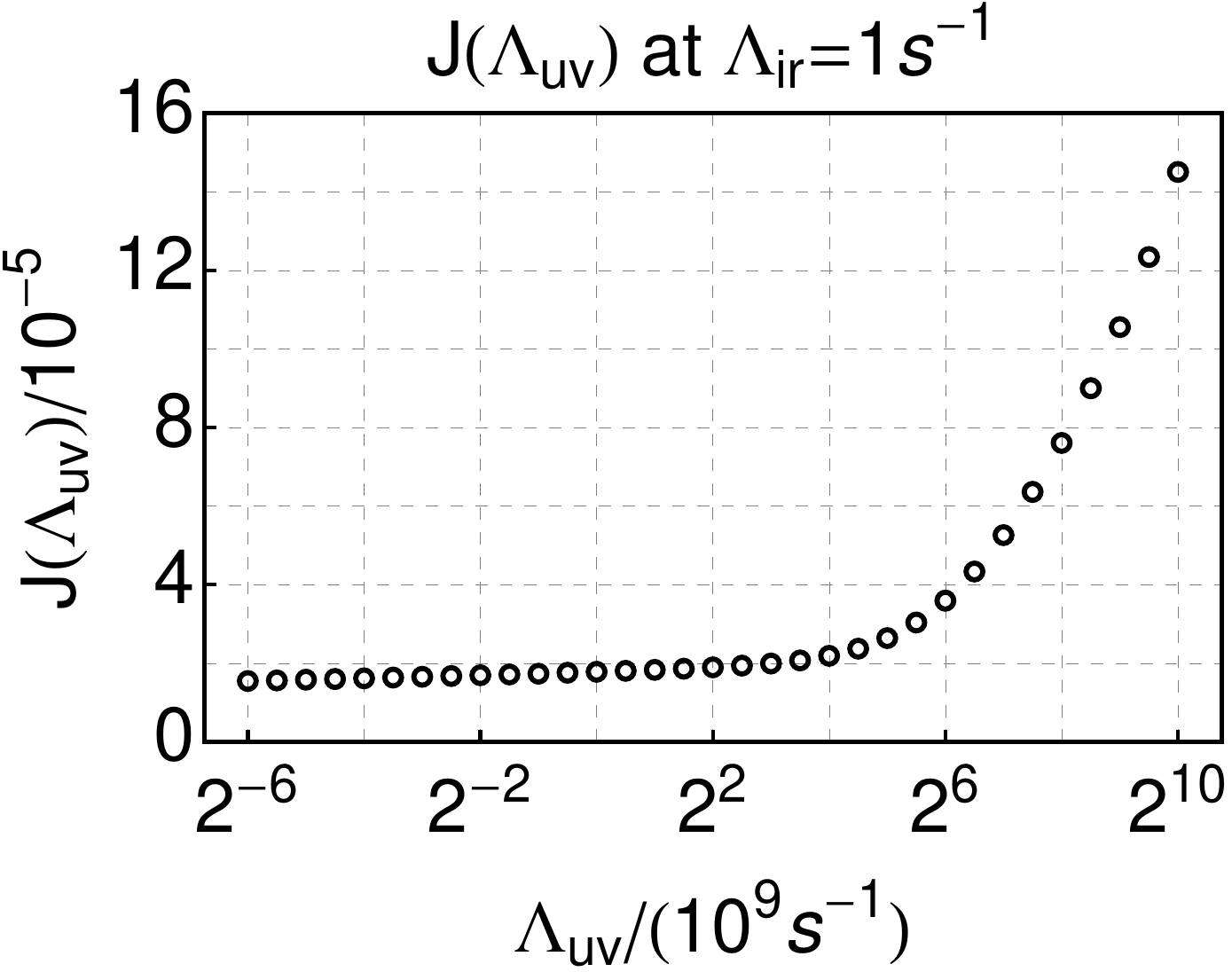}
\end{tabular*}
\caption{\label{co}(Color online) (a) Variation in the cost functional with respect to changes in the infrared and ultraviolet cutoff frequency. We used the pulse sequences shown in Figs.~\ref{pt}(b)~and~\ref{pt}(c). Inset, solid line: Noise spectrum employed for our calculations. Dashed and dotted--dashed lines represent $1/f$ and $f$ noise contributions, respectively. (b)~For a given ultraviolet cutoff, the cost functional shows a logarithmic dependence on $\Lambda_{ir}$, \textit{i.e.} $J(\Lambda_{ir})\sim \log{\left( 1/\Lambda_{ir}\right) }$, as expected (see Ref.~\onlinecite{Makh04}). (c)~For $J(\Lambda_{uv})$ we obtain a logarithmic dependence over a broad regime. However, when Ohmic contributions become relevant the cost functional increases significantly.}
\end{figure}

Our strategy is to use short control pulses of high amplitude which move the qubit away from the degeneracy point only briefly and are within current experimental capabilities. 
We employ a time--optimal control strategy which incorporates the final time of the gate operation~($t_f$) as an additional control parameter, compatible with experimental control field strength.  
The control fields are of the form,
\begin{eqnarray}
n_c(t)&=&\frac{1}{2}+g(t)A_c\sin{\left( \omega_c t+ \varphi_c \right) } e^{ -\gamma_c\left( t-t_c^0\right)^2}, \nonumber \\
\frac{\Phi(t)}{\Phi_0}&=&g(t)\left\lbrace \frac{1}{2}- \frac{A_\Phi}{4}\left[\left\lbrace 1 + \sin{\left( \omega_\Phi t+ \varphi_\Phi \right) }\right\rbrace \times \right. \right. \nonumber \\
&&  \left. \left.   e^{ -\gamma_\Phi\left( t-t_\Phi^0\right)^2}\right]\right\rbrace ,  \nonumber 
\end{eqnarray}
with $\left\lbrace A_{c/\Phi},\omega_{c/\Phi},\varphi_{c/\Phi},\gamma_{c/\Phi},t_{c/\Phi}^0\right\rbrace$ representing free parameters.
Additionally, we restrict $\left\lbrace n_c(t), \Phi(t)\right\rbrace $ to start and end at the degeneracy point $(n_c=1/2,\Phi=0)$ by using an envelope function $g(t)$ [dashed, gray line in Fig.~\ref{pt}(b)] and to satisfy other constraints imposed by the design of the superconducting circuit. 
There are other methods, such as  driving the qubit by NMR--like techniques and adiabatic pulses, within which fidelities of $\sim 0.3-0.4$ have been achieved.~\cite{Collin04}  Gate fidelities of $\sim 0.4$ have also been reported for a CNOT gate implemented by two coupled flux qubits.~\cite{Plant07} Recently, single qubit operations with gate errors of $1\sim2\%$ have been realized within phase and transmon qubits.~\cite{Lucero08,Chow09}

To optimize the cost functional Eq.~\eqref{cost} we use a parallelized, constrained version of a differential evolution algorithm with 300 individuals per generation and 3000 generations per optimization run.~\cite{DEA} The algorithm finds an optimum for control fields given in Figs.~\ref{pt}(b)~and~\ref{pt}(c). We typically choose crossing probabilities ($C$) and scaling parameters ($S$) within a range of $C \in [0.90,0.96]$ and $S \in [0.75,0.85]$. Variations in these parameters have a large impact on the convergence rate of the algorithm. All calculations have been performed on a 40--node SUN Linux Cluster.

For $t_f\approx 50$~ps,  gate fidelity ($F$) losses  $\Delta F= 1-F\equiv \left( J[\epsilon^*]/J_{max}\right)^{1/2} $ as low as $\approx 10^{-3}$ are achievable.  A plot of the associated matrix--element deviations from the Hadamard operation, $\left|\left( \hat \chi- \hat \chi^D\right)_{mn}\right|$, is shown in Fig.~\ref{pt}(a).  Losses from decoherence (mainly included in red/dark gray bars) dominate over those from leakage (mainly included in white bars) by about a factor of $15$, largely due to Ohmic contributions in the noise spectrum. 
To demonstrate the importance of including additional Cooper pair occupation on the superconducting island within the model, we optimized the gate without leakage and then used these control fields to steer the qubit subjected to leakage. The value of the cost functional increases by about 4 orders of magnitude corresponding to an increase in $\Delta F$ by 2 orders of magnitude. Incorporation of leakage states $\ket{-2}$ and $\ket{3}$ does not add significant fidelity losses for optimal pulses given in Figs.~\ref{pt}(a)~and~\ref{pt}(b). Therefore, expanding the leakage state--space beyond $\ket{-1}$ and $\ket{2}$ is not necessary.

We have also explored  longer pulses ($t_f=200\einheit{ps}$)  and do not start at the degeneracy point with respect to $\Phi$ [the working point is now set at $(n_c=1/2, \Phi=\Phi_0/2)$, see Figs.~\ref{pt}(d)-\ref{pt}(f)]. These pulses show an enhanced reduction of leakage but they do lead to stronger decoherence because of an increased pulse--duration. Also, moving the qubit's working point away from the degeneracy point causes higher sensitivity to fluctuations in the control fields.  Typical fidelity losses $\Delta F$  for $200\einheit{ps}$ pulses are about 6 times higher than those for $50\einheit{ps}$ ones ($\Delta F \approx 10^{-2}$).\par

Time--optimal control inevitably leads to short pulses (typically of the order of $50\einheit{ps}$ for present field intensities)  whose generation in experiment is a demanding task. However, picosecond electrical pulses can be produced e.g. by optoelectronic devices, such as photocunductive switches~\cite{Holz00} or by optical rectification of ultrashort optical pulses using nonlinear media (e.g. LiTaO$_3$).~\cite{Naha98} Ultrafast pulse--shaping methods have been discussed extensively in Ref.~\onlinecite{Spoerl07}.\par

\section{Conclusion}
By applying a newly developed, process--tomography--based optimal control theory for open quantum systems to a Josephson charge qubit, we have shown that one--qubit gates, such as the Hadamard gate, can be realized with remarkably high fidelity. The strategy has been to keep deviations of the control fields with respect to the degeneracy point as short as possible while performing the desired unitary operation in time--optimized fashion compatible with experimentally available control field strength. A fully quantum mechanical description based on experimental noise spectra has been employed to model dephasing effects and additional non--computational basis states have been included to account for unwanted Cooper--pair occupation on the superconducting island. Depending on the gate operation time, which has been treated as a variable, we could achieve fidelities of the order of $F\approx 1-10^{-3}$ to $1-10^{-2}$. In terms of the process tomography matrix this corresponds to errors of the order $\operatorname{Abs}\left[ \left ( \hat \chi - \hat\chi^D\right)_{ mn}\right] \approx 10^{-3}-10^{-2}$. Thus we find that charge qubits can be made to perform at an equal level with current realizations of other Josephson qubits (such as phase, flux or transmon qubits) which have been tested in experiment so far.

\section{Acknowledgment}
The authors wish to acknowledge financial support of this work by FWF under Project No. P18829, as well as helpful discussions with M. Wenin.

\end{document}